\documentclass[aps,prl,superscriptaddress,showpacs,preprint]{revtex4}
\usepackage{graphicx}
\begin{document}

\title{Configuration interaction method for Fock-Darwin states}

\author{Andreas Wensauer}
\affiliation{Institute for Microstructural Sciences, National Research
         Council of Canada, Ottawa, Canada, K1A 0R6}
\affiliation{Institut f\"ur Theoretische Physik, 
         Universit\"at Regensburg, D-93051 Regensburg, Germany}

\author{Marek Korkusi\'nski}
\affiliation{Institute for Microstructural Sciences, National Research
         Council of Canada, Ottawa, Canada, K1A 0R6}

\author{Pawel Hawrylak}
\affiliation{Institute for Microstructural Sciences, National Research
         Council of Canada, Ottawa, Canada, K1A 0R6}

\begin{abstract}
We present a configuration interaction method optimized for
Fock-Darwin states of two-dimensional quantum dots with an axially
symmetric, parabolic confinement potential subject to a perpendicular
magnetic field. 
The optimization explicitly accounts for  geometrical and dynamical
symmetries of the Fock-Darwin single-particle states and for many-particle
symmetries associated with the center-of-mass motion and with the total spin.
This results in a basis set of reduced size and improved accuracy. 
The numerical results compare well with the quantum Monte Carlo and
stochastic variational methods. 
The method is illustrated by the evolution of a strongly correlated
few-electron droplet in a magnetic field in the regime of the fractional quantum Hall effect.
\end{abstract}

\pacs{71.15.-m (Methods of electronic structure calculations),
  73.21.La (Quantum dots),
  73.23.Hk (Coulomb blockade, single electron tunnelling)}

\maketitle

\section{Introduction}
In the configuration interaction (CI) method the Hamiltonian of an 
interacting system is calculated in the basis of many-electron
configurations and diagonalized exactly.  
Unlike the correlated basis of Jastrow or Hylleraas functions, the CI
basis functions do not automatically include correlations among pairs
of electrons, so their repulsive interaction is not minimized. 
Hence the size and choice of the basis influences the accuracy of
results. 
Exact diagonalization (ED) has been an important tool used to
investigate the electronic and optical properties of quantum 
dots \cite{jacak98,maksym90,merkt91,pfannkuche93,hawrylak93,yang93,hawrylak93b,palacios94,wojs95,oaknin96,wojs96,maksym96,hawrylak96,wojs97,wojs97b,eto97,eto97b,maksym98,imamura98,imamura98b,hawrylak99,creffield99,bruce00,reimann00,mikhailov02}.
Not only does it provide benchmark results for the ground state
energy and wave function, but also it gives access to excited states 
and hence permits an interpretation of a number of spectroscopic
techniques such as  far-infrared \cite{wojs96},
photoluminescence \cite{wojs97}, and Raman \cite{wojs97b}.
In contrast, the quantum Monte Carlo (QMC) \cite{bolton96,harju99,harju99b,harju02a,harju02b,egger99,pederiva00,pederiva03,filinov01},
mean-field Hartree-Fock \cite{yannouleas99,reusch01},
and density-functional theory (DFT) calculations
\cite{koskinen97,austing99,hirose99,steffens99,wensauer00,wensauer01} 
are restricted to the ground-state properties.

The objective of this work is to present a CI method optimized for the
specific problem of interacting electrons in a parabolic dot in a
magnetic field. 
The optimization explicitly accounts for  geometrical and dynamical
symmetries of  single-particle states, and for many-particle
symmetries associated with the center-of-mass (CM) motion and total
spin. 
This results in a reduced basis size and improved accuracy,
which in turn allows for a reliable computation of dot properties in
the strongly correlated regime. 
Previous work on total spin and CM-resolved calculations used either the
group-theoretical approach limited to a small number of
electrons \cite{hawrylak93,hawrylak93b}, or the lowest Landau level
approximation \cite{oaknin96}.
Here we build on these results to construct a reliable and versatile
computational tool capable of testing many of the interesting
predictions involving electronic correlations in quantum dots. 
The paper is organized as follows.
In Sec. II we focus on the physical properties of our system,
discuss the Hamiltonian, its symmetries and quantum numbers, and its
scaling properties. 
In Sec. III we present a total spin- and CM-resolved CI technique. 
We compare our results with QMC and stochastic-variational method
(SVM) and with previous results describing the evolution of the
electronic droplet with the magnetic field. 

\section{Hamiltonian and symmetries}

We consider a two-dimensional quantum dot with a
parabolic confinement potential of strength $\omega_0$ (henceforth
$\hbar=1$) in a perpendicular magnetic field ${\bf B}=(0,0,B)$.
The Hamiltonian for $N$ particles in real-space representation [with
${\bf r}=(x,y)$, $\hat {\bf p}= (\hat p_x,\hat p_y)$, and the vector
potential  ${\bf  A}({\bf r})=B/2(-y,x)$] is: 
\begin{eqnarray}
\hat H=\sum_{j=1}^N \left(\frac{1}{2m^\ast}\left(\hat{\bf p}_j+
e{\bf A}({\bf r_j}) \right)^2 +
\frac{1}{2}m^\ast\omega_0^2{\bf r}_j^2 \right)
+\frac{1}{2} \sum_{j,k=1}^{N}{\!\! ^\prime}\,\,
\frac{e^2}{4\pi\varepsilon\varepsilon_0|{\bf r}_j-{\bf r}_k|},
\label{ham_rss}
\end{eqnarray}
with the Zeeman term omitted for simplicity.
Here $m^\ast$, $e$ are the electronic effective mass and charge,
respectively, $\omega_c=eB/m^\ast$ is the cyclotron energy, 
and $\varepsilon$ is the dielectric constant of the host semiconductor.
We can rewrite the one-particle contribution of equation (\ref{ham_rss})
to obtain
$ \left({\bf p}_j^2/2m^\ast + m^\ast\omega_h^2{\bf r}_j^2 / 2+
(\omega_c/2)\hat l_j\right)$, 
where $\omega_h=\sqrt{\omega_0^2+\omega_c^2/4}$ is the hybrid
frequency and $\hat l_j$ is the angular momentum operator.  
Introducing the effective length 
$l_{\rm h}^{-2}=m^\ast \omega_{\rm h}$, and scaling all lengths in
the units of $l_{h}$ and energies in effective Rydbergs 
$Ry = m^*e^4/2\varepsilon^2$ results in a
dimensionless Hamiltonian,  
\begin{eqnarray}
\hat H=   \frac{1}{(l_{h}/a_0)^2}
\left( \sum_{j=1}^N   \left( -{ \bf \nabla }_j^2+ {\bf r}_j^2
+\frac{l_{\rm h}^2}{l_{0}^2} \hat l_j     \right)
+\left( {l_{\rm h}/a_0} \right) \sum_{j,k=1}^{N}{\!\! ^\prime}\,\,
|{\bf r}_j-{\bf r}_k|^{-1} \right).
\end{eqnarray}
Here  the strength of  Coulomb interactions is proportional to the
effective length $(l_{\rm h}/a_0 )$, with 
$a_0=\varepsilon/m^*e^2$ being the effective
Bohr radius and $l_0^{-2}=m^\ast\omega_{\rm c}$ being the magnetic
length.
Therefore, the larger the effective length or the smaller the hybrid
frequency, the larger the contribution from electron-electron
interactions. 

The Hamiltonian (\ref{ham_rss}) exhibits another scaling property.
To see it we rewrite it as a function of $\omega_0$,
$\omega_c$, and $N$ and separate the contribution of the angular
momentum. 
This concept is similar to the modification of the Hamiltonian of a
(natural) atom in Ref.~\cite{capelle02}.
\begin{equation}
\hat H=\sum_{j=1}^N \left(\frac{\hat{\bf p}_j^2 }{2m^\ast}+
\left(\omega_0^2+\frac{\omega_c^2}{4}\right)
\frac{m^\ast{\bf  r}_j^2}{2}\right)
+ \sum_{j,k=1}^{N} {\!\! ^\prime}\,\,
\frac{e^2}{8\pi\varepsilon\varepsilon_0|{\bf r}_j-{\bf r}_k|}
+\sum_{j=1}^{N}\frac{\omega_c}{2}\hat l_j .
\label{scaling2}
\end{equation}
The total Hamiltonian is therefore a sum of a Hamiltonian of a dot in
zero magnetic field with the confinement frequency
$\sqrt{\omega_0^2+\omega_c^2/4}$, and the total angular momentum
contribution $(\omega_c/2)\hat L$. 
As a result we can map the spectra and wave functions of a dot in the
presence of the magnetic field to those at zero field. 
Note that Eq.~(\ref{scaling2}) implies that the energy levels of
subspaces with different angular momenta are shifted against each
other due to the term $(\omega_c/2)\hat L$, whereas within a specific
subspace of the angular momentum the energy levels are fixed relative
to each other.  
Therefore, ground states in high magnetic fields are excited states
in zero field and a suitable choice of a finite basis is not trivial.

The choice of the basis is facilitated by the properties of the one-
and the many-particle Hamiltonian. 
The properties of the one-electron Hamiltonian are easily understood 
by a transformation from the momentum and position representation 
($\hat{\bf p},{\bf r}$) into the language of harmonic-oscillator
raising and lowering operators \cite{hawrylak93,hawrylak93b}:
\begin{equation}
\hat H_0=\omega_+(\hat a_+^\dag\hat a_++1/2)+
\omega_-(\hat a_-^\dag\hat a_-+1/2),
\end{equation}
with the energy spectrum
$\varepsilon_{nm\sigma}=\omega_+(n+1/2)+\omega_-(m+1/2)$
($n,m=0,1, ..., \infty$), and the two oscillator frequencies
$\omega_\pm=\omega_h\pm\omega_c/2$. 
The harmonic-oscillator spectrum is equivalent to the spectrum derived
independently by Fock \cite{fock28} and Darwin \cite{darwin31}
(the Fock-Darwin spectrum, FD) applying different methods. 
The corresponding states are denoted by $|nm\sigma\rangle$, where
$\sigma=\uparrow,\downarrow$ denotes the electronic spin.
The eigenvalues of the angular momentum operator $\hat l=\hat n-\hat
m$, the spin $\hat {\bf s}^2$, and its  $z$-component $\hat s_z$ are
good quantum numbers. 
In addition, there are dynamical symmetries associated with the
form of the confining potential. 
These symmetries lead to degeneracies of the harmonic-oscillator
spectrum at $B=0$ and also at finite fields whenever $\omega_+/\omega_-=k$,
with $k=1,2,...$. 

For numerical calculations it is useful to resort to the language of
second quantization. 
The Hamiltonian of $N$ particles in second quantization with Fermion
creation $\hat c^\dag_{nm\sigma}$ and annihilation $\hat c_{nm\sigma}$
operators can be written as: 
\begin{equation}
\hat H=
\sum_{i\sigma}\varepsilon_{i\sigma} \hat c^\dag_{i\sigma}\hat c_{i\sigma}
+{1\over 2} \sum_{ijkl\sigma\sigma^\prime}
\langle i,j|\hat W|k,l\rangle \nonumber\\
 \hat c^\dag_{i\sigma}\hat c^\dag_{j\sigma^\prime}
\hat c_{k\sigma^\prime}\hat c_{l\sigma},
\label{ham_2q}
\end{equation}
where the composite index $i=(n,m)$ (similarly $j,k,l$) denotes the
harmonic-oscillator quantum numbers.
The Coulomb matrix elements $\langle |\hat W|\rangle$ are given
explicitely in Ref.~\cite{hawrylak_sc93}.

The Hamiltonian (\ref{ham_2q}) conserves total spin,
total z-component of spin, $\hat S_z=1/2\sum_{i\sigma}\sigma
\hat c^\dag_{i\sigma}\hat c_{i\sigma}$, 
and total angular momentum, $\hat L=\sum_{nm\sigma}(n-m)
\hat c^\dag_{nm\sigma}\hat c_{nm\sigma}$.
Apart from the spin and the geometrical and dynamical symmetries
discussed above, the many particle Hamiltonian (\ref{ham_2q}) has also
a hidden symmetry related to the separability of the CM and relative
motion \cite{hawrylak93b}.
The CM symmetry  can be expressed by the introduction of new
operators describing collective CM excitations of the system:
\begin{eqnarray}
\hat A_+^\dag &=&{1\over\sqrt{N}}
\sum_{nm\sigma}\sqrt{n+1}\,
\hat c^\dag_{(n+1)m\sigma}\hat c_{nm\sigma},\nonumber\\
\hat A_-^\dag&=&{1\over\sqrt{N}}
\sum_{nm\sigma}\sqrt{m+1}\,
\hat c^\dag_{n(m+1)\sigma}\hat c_{nm\sigma},\nonumber\\
\hat A_+&=&{1\over\sqrt{N}}
\sum_{nm\sigma}\sqrt{n}\,
\hat c^\dag_{(n-1)m\sigma}\hat c_{nm\sigma},\nonumber\\
\hat A_-&=&{1\over\sqrt{N}}
\sum_{nm\sigma}\sqrt{m}\,
\hat c^\dag_{n(m-1)\sigma}\hat c_{nm\sigma}.
\end{eqnarray}
The operators $\hat A_+^\dag$ and $\hat A_-^\dag$ ($\hat A_+$ and
$\hat A_-$) satisfy the following commutation relation with the 
full interacting Hamiltonian $\hat H$: 
$[\hat H,\hat A_\pm^\dag]= \omega_\pm\hat A_\pm^\dag $ and
$[\hat H,\hat A_\pm]=-\omega_\pm\hat A_\pm$.
They can be interpreted as creation (annihilation) operators for
collective excitations of the system since they increase (decrease)
the energy of the system by $\omega_\pm$.
They allow for the construction of exact eigenstates of the
interacting system.  

We can now explicitly define CM operators $C$ which commute with
the interacting Hamiltonian: 
\begin{eqnarray}
\hat C_+&=&\hat A_+^\dag\hat A_+=
{1\over N}\sum_{n^\prime m^\prime\sigma^\prime nm\sigma}
\sqrt{(n^\prime+1)n}\,
\hat c^\dag_{(n^\prime+1)m^\prime\sigma^\prime}
\hat c_{n^\prime m^\prime\sigma^\prime}
\hat c^\dag_{(n-1)m\sigma}\hat c_{nm\sigma},\nonumber\\
\hat C_-&=&\hat A_-^\dag\hat A_-=
{1\over N}\sum_{n^\prime m^\prime\sigma^\prime nm\sigma}
\sqrt{(m^\prime+1)m}\,
\hat c^\dag_{n^\prime(m^\prime+1)\sigma^\prime}
\hat c_{n^\prime m^\prime\sigma^\prime}
\hat c^\dag_{(m-1)m\sigma}\hat c_{nm\sigma}.
\label{cmoperators}
\end{eqnarray}

Because Slater determinants constructed from a FD  basis set are
eigenstates of $\hat L$ and $\hat S_z$, it is natural to resolve the
total angular momentum $L$ and the total spin in $z$-direction $S_z$
as good quantum numbers. 
On the other hand, the structure of the operators $\hat {\bf S}^2$ 
and $\hat C_\pm$ in second quantization is similar to that of the
Coulomb interaction, involving two creation and two annihilation
operators.
In what follows we construct ED methods which make use of these
operators  to reduce the size of the Hamiltonian matrix. 

\section{Spin-resolved CI method }

The form of the total spin operator in the language of creation and
annihilation operators is:
\begin{equation}
\hat {\bf S}^2=
N/2+\hat S_z^2-\sum_{ij}
\hat c^\dag_{j \uparrow}\hat c^\dag_{i\downarrow} 
\hat c_{j\downarrow}\hat c_{i\uparrow}.
\end{equation}
Its last term annihilates one particle in a state $i$ and
recreates it on the same orbital, but with flipped spin;
the same is true for the orbital $j$. 
Thus, the pattern of orbital quantum numbers is conserved, which makes
it possible to arrange the Slater determinants in blocks which enclose
all states with the same pattern, and to rotate the single blocks
into eigenstates of $\hat {\bf S}^2$.
The size of the blocks can be obtained using combinatoric arguments.
We start with a given electron number $N$, among which $N_\uparrow$
electrons are spin up, and $N_\downarrow$ spin down, and we distribute
them on $M$ orbitals ($M\leq N$). 
Because of Pauli's exclusion principle we get
\begin{equation}
{N\over2}\le\max{(N_\uparrow,N_\downarrow)}\le M\le N.
\end{equation} 
Thus we have $N-M$ doubly occupied orbitals (singlets), $M-N_\downarrow$
unpaired spins up, and $M-N_\uparrow$ unpaired spins down.
The distribution of the singlets is determined by Pauli's principle.
The number of different orbitals available for unpaired spins up and
unpaired spins down is $M-(N-M)$. 
The size of the block equals the number of all possible distributions
of unpaired electrons spin up and down on $M-(N-M)$ orbitals:
\begin{eqnarray}
{\rm blocksize} &=& {M-(N-M) \choose M-N_\downarrow}
{M-(N-M)-(M-N_\downarrow) \choose M-N_\uparrow} \nonumber\\
&=& \frac{(2M-N)!}{(M-N_\downarrow)!(M-N_\uparrow)!}.
\end{eqnarray}
In Table \ref{block1} we show the number of total spin eigenstates per
block for a six-electron system as a function of $M$ and
$S_z=(N_\uparrow-N_\downarrow)/2$.  
These blocks are fairly small (e.g., ${\rm blocksize}=252$ for
$N=M=10$, $S_z=0$) and the diagonalization  of the blocks has to be
done only once for each value of $M$. 

In Table \ref{matrix_red} we show how the size of the Hamiltonian is
reduced by applying the scheme on a system with $N=5$, $L=1$, 
$S_z=1/2$ and $352954$ Slater determinants and on a system with
$N=6$, $L=S_z=0$ and $326120$ Slater determinants. 
Note that for even electron numbers $N$ the largest subspace is that
of $S=1$, not $S=0$. 
In the case of odd $N$ the $S=1/2$ subspace is larger than the others.
\begin{table}
{\centering
\begin{tabular}{c|c|c|c|c}
\hline
            & M=3    & M=4    & M=5    & M=6    \\ \hline
$S_z=0$     & $S=0$: 1 
            & $S=1$: 1
            & $S=0$: 2, $S=1$: 3, 
            & $S=0$: 5, $S=1$: 9, \\
          &&& $S=2$: 1 
            & $S=2$: 5, $S=3$: 1 \\ \hline
$S_z=\pm 1$ && $S=1$: 1
             & $S=1$: 3, $S=2$: 1 
             & $S=1$: 9, $S=2$: 5, \\
             &&&& $S=3$: 1 \\ \hline
$S_z=\pm 2$ & & & $S=2$: 1 & $S=2$: 5, $S=3$: 1\\ \hline
$S_z=\pm 3$ & & & & $S=3$: 1\\ \hline
\end{tabular}\\}
\caption{Number of total spin eigenstates per block for a six-electron
  system as a function of $M$ and $S_z$. 
\label{block1}}
\end{table}
\begin{table}
{\centering
\begin{tabular}{c|c|c|c|c}
\hline
$N$  & S=1/2  & S=3/2  & S=5/2 & $\Sigma$  \\ \hline
$5$  & 184769 & 137705 & 30480 & 352954 \\ \hline
\end{tabular}\vspace{1mm}\\
\begin{tabular}{c|c|c|c|c|c}
\hline
$N$  & $S=0$ & $S=1$  & $S=2$ & $S=3$ & $\Sigma$ \\ \hline
$6$  & 92410 & 152460 & 70711 & 10539 & 326120\\ \hline
\end{tabular}\\}
\caption{Effective size of the Hamiltonian for systems with $N=5$,
$L=1$, $S_z=1/2$ and 352954 Slater determinants and with $N=6$,
$L=S_z=0$ and 326120 Slater determinants.
\label{matrix_red}}
\end{table}

We now apply the spin-resolved CI method to the calculation
of  the ground state  energy of 4 through 7 electrons. 
The accuracy of the method is assessed by comparison of our
calculations in zero magnetic field with available results from
quantum Monte Carlo and stochastic-variational method.
The calculations were performed using GaAs parameters
($\varepsilon=12.4$, $m^\ast=0.067$) and a confinement energy
$\omega_0=3.32$ meV. 
Table \ref{etable} shows the ground state energies for a number of
subspaces $L$, $S$. 
We see that the QMC and SVM energies appear to be slightly lower then
the ED results, however, their order appears to change depending on
the actual subspace  and number of electrons. 
Altogether, the ED appears to give highly reliable results for ground
and excited states which can be extended to finite magnetic fields. 
\begin{table}
{\centering
\begin{tabular}{c|c|c|c|c|c}
\hline
{$N$} & {$(L,S)$}   & $E_{\rm QMC}[{\rm meV}]$  & $E_{\rm SVM}[{\rm
meV}]$ & $E_{\rm ED}[{\rm meV}]$ & {rel. error of ED}\\ \hline
4     &  $(0,1)$    & 44.0439   & 44.0261 & 44.0763 & $<1.2\cdot 10^{-3}$\\ 
\hline 
4     &  $(-2,0)$   & 44.5301   & 44.4945 & 44.5498 & $<1.3\cdot 10^{-3}$ \\ 
\hline
4     &  $(0,0)$    & 44.8301 & 44.8004 & 44.8355 & $<8\cdot 10^{-4}$\\ 
\hline
5     &  $(-1,1/2)$ & 65.6159   & 65.5827 & 65.6488 & $<10^{-3}$  \\ \hline 
6     &  $(0,0)$    & 90.1166   & 90.1391 & 90.2383 & $<1.5\cdot 10^{-3}$\\ \hline 
7     &  $(-2,1/2)$    & 118.9784   & --- & 119.1426 & $<1.2\cdot 10^{-3}$\\ \hline 
7     &  $(0,1/2)$    & 119.3045   & --- & 119.4069 & $<1.0\cdot 10^{-3}$\\ 
\hline
\end{tabular}\\}
\caption[]{Comparison of ground state energies for $N=4$, 5, 6, 7 of the
subspaces $L,S$ ($\omega=3.32$ meV) at $B=0$. 
QMC energies are from Ref.~\cite{pederiva03}, and
SVM energies from Ref.~\cite{varga01}. 
SVM results for seven electrons are not available.
\label{etable}}
\end{table}

\section{Spin- and CM-resolved CI method}

The ED method with both the total spin and the CM resolved as good
quantum numbers is particularly applicable in the high magnetic field
strongly correlated regime, where correlation effects dominate the
reconstruction of the dot through magic states
\cite{yang93,imamura98b,chakra92}, spin textures \cite{hawrylak99}, 
and skyrmions \cite{oaknin96}.

The construction starts with defining the number of electrons $N$ and
their total angular momentum $L$.
Moreover, we restrict the single-particle basis set by choosing the
number of Landau levels of interest. 
This is done by fixing the maximal quantum number $n_{\rm max}$.
Since the total angular momentum is defined using the FD numbers $n$
and $m$, the above choices impose a limitation on the number $m$,
i.e., determine  an $m_{\rm max}$ such that
$\sum_{j=1}^N m_j=\sum_{j=1}^N n_j-L
\le  n_{\rm max}N-L\le m_{\rm max}$. 
The Hilbert space is thus finite and we can generate all
configurations and diagonalize them with respect to the CM and total
spin operators.  
Since we restrict ourselves to $n_{\rm max}$ Landau levels, only the
CM operator $\hat C_-$ is of interest. 
From its definition (Eq.~(\ref{cmoperators})) we see
that $\hat C_-$ increases/decreases the quantum number $m$, but does
not change the pattern of numbers  $n\sigma$.
We can therefore organize our configurations in blocks with the same
$n\sigma$ pattern; these blocks are diagonalized separately.
The general scheme is thus similar to that for the total spin, only
here the CM diagonalization has to be done for each block as the
matrix elements of $\hat C_-$ in second quantization explicitly
depend on $m$.
We retain only the eigenstates with eigenvalue $C_-=0$, because
the eigenstates corresponding to eigenvalues $1,2,3,..$ 
can be generated systematically by applying the CM motion
creation operator $\hat A_-^\dag$ on eigenstates with eigenvalue $0$. 
 
\begin{figure}[h]
\includegraphics*[width=0.9\textwidth]{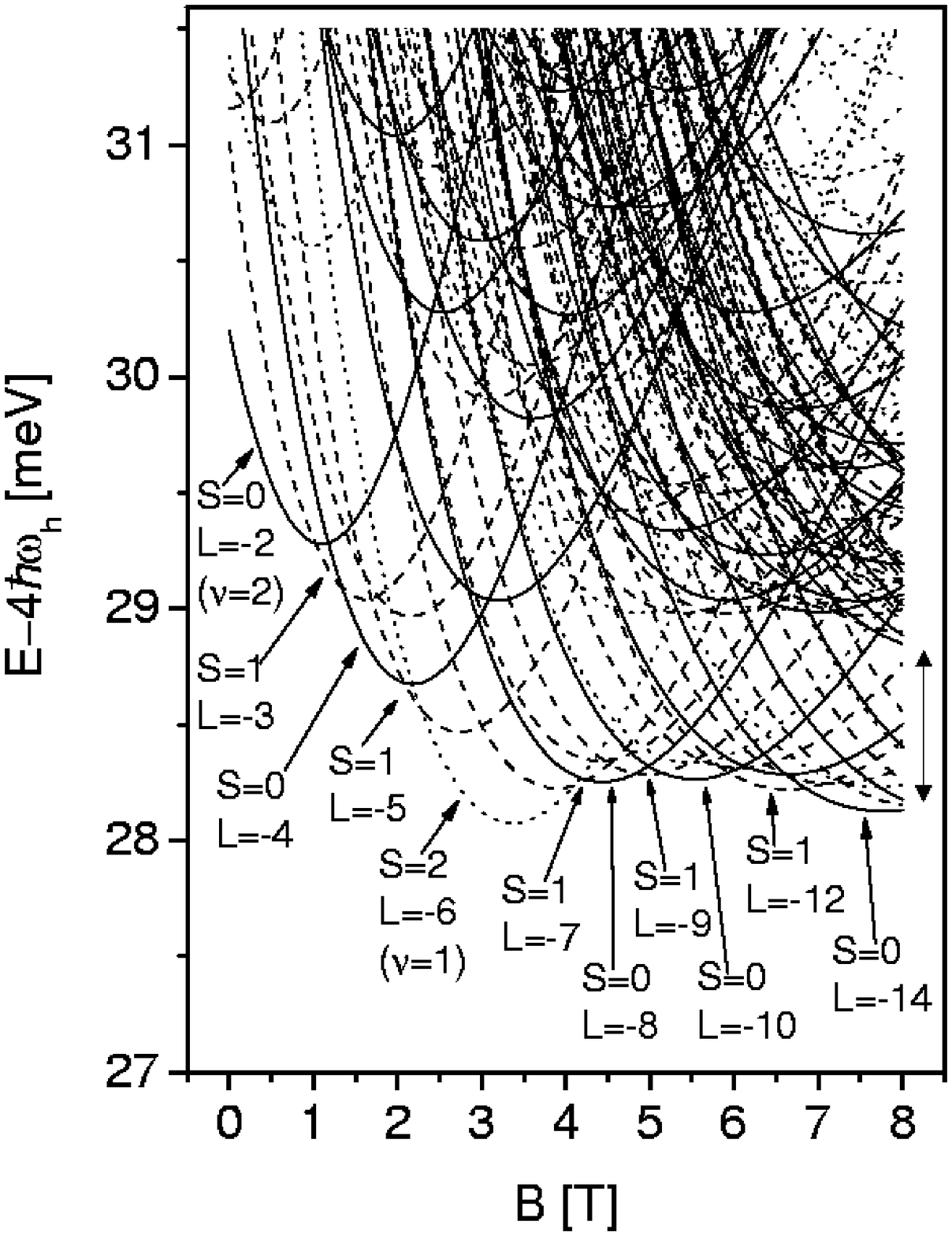}
\caption{
Energies of the ground state and the two lowest-lying excited states
of four electrons with angular momenta $L=-25$ to $0$ calculated in
the two-Landau-level approximation. 
The confinement energy is $\omega=3$ meV and the Zeeman energy is
neglected.
All these states are eigenstates of the center-of-mass operator with
the quantum number $C_-=0$;
solid lines correspond to total spin for $S=0$, dashed lines - to
$S=1$, and dotted lines to $S=2$.
\label{figcm}
}
\end{figure}

Our next aim is to resolve the total spin in the basis of stored CM
eigenvectors. 
We have established that $\hat {\bf S}^2$ couples states with the same
$nm$ pattern. 
It is therefore convenient to introduce superblocks with the
same $n$ pattern. 
Thus, a superblock encloses all blocks with the same $n\sigma$ pattern
and $\hat {\bf S}^2$ can couple states only within a superblock. 
After calculating all eigenstates with zero CM motion in the previous
step, all states with the same $n$ pattern are collected and
diagonalized with respect to the total spin operator $\hat{\bf S}^2$. 
Note that one set of CM eigenvectors calculated for the system with
lowest possible spin polarization ($0$ for even number of electrons,
$1/2$ for odd number of electrons) can be used to calculate all
possible eigenvalues of $\hat {\bf S}^2$.
As a result we obtain a basis set consisting of
eigenstates of the operators $\hat L$, $\hat S_z$, $\hat {\bf S}^2$,
and $\hat C_-$, which we use to generate the Hamiltonian matrix.
Table \ref{tablecm} shows how the Hilbert space is reduced for the
$(S_z,L)=(0,-18)$ subspace of a four-electron system in the
two-Landau-level approximation ($n_{\rm max}=1$).  
The first column labels the superblocks. 
The second column contains the information which electrons are in the
lowest ($n=0$), and which are in the second Landau level ($n=1$), and
how many configurations are possible (third column). 
Each superblock may enclose several blocks which are labelled in the
fourth column. 
The $n\sigma$-pattern and the corresponding number of configurations
are in the fifth and sixth columns. 
After applying the CM operator and filtering out states with zero CM
motion the number of states is reduced for each $n\sigma$ block
(seventh column), and to $1555$ in total (see last row). 
In the next step all superblocks are diagonalized with respect to the
total spin operator. 
The resulting numbers of configurations for $S=0$, 1, and 2 are given
in the last column. 
Thus, instead of having to solve an eigenvalue problem of the size
$15456$ in the subspace $S_z=0$, $L=-25$ we end up with three
eigenvalue problems with matrix sizes $528$, $791$, and $236$ (see
last row). 
\begin{table}
{\centering
\begin{tabular}{c|c|c|c|c|c|c|c}
\hline
Superblock & ${n}$ pattern & Size & Block & ${n\sigma}$ pattern &
Size & $C_-=0$ & Spin resolved \\ \hline
1 & $(0,0,0,0)$ & 728  & a &
$(0\downarrow,0\downarrow,0\uparrow,0\uparrow)$ & 728 & 78 &
$S=0$: 26 \\
  &&&&&&& $S=1$: 42 \\
  &&&&&&& $S=2$: 10 \\ \hline
2 & $(0,0,0,1)$ & 3458 & a &
$(0\downarrow,0\downarrow,0\uparrow,1\uparrow)$ & 1729 & 182 &
$S=0$: 126 \\
  & & & b & $(0\downarrow,1\downarrow,0\uparrow,0\uparrow)$ &
  1729 & 182 & $S=1$: 182 \\
  &&&&&&& $S=3$: 56 \\ \hline
3 & $(0,0,1,1)$ & 5880 & a &
$(0\downarrow,0\downarrow,1\uparrow,1\uparrow)$ & 910 & 91 &
$S=0$: 196 \\
  & & & b &
$(0\downarrow,1\downarrow,0\uparrow,1\uparrow)$ & 5880 & 406 &
$S=1$: 301 \\
  & & & c &
$(1\downarrow,1\downarrow,0\uparrow,0\uparrow)$ & 910 & 91 &
$S=2$: 91 \\ \hline
4 & $(0,1,1,1)$ & 4270 & a &
$(0\downarrow,1\downarrow,1\uparrow,1\uparrow)$ & 2135 & 210 &
$S=0$: 145 \\
  & & & b &
$(1\downarrow,1\downarrow,0\uparrow,1\uparrow)$ & 2135 & 210 &
$S=1$: 210 \\
  &&&&&&& $S=2$: 65 \\ \hline
5 & $(1,1,1,1)$ & 1120 & a &
$(1\downarrow,1\downarrow,1\uparrow,1\uparrow)$ & 1120 & 105 &
$S=0$: 35 \\
  &&&&&&& $S=1$: 56 \\
  &&&&&&& $S=2$: 14 \\ \hline
$\Sigma$ & & 15456 &&& 15456 & 1555 & $S=0$: 528\\
  &&&&&&& $S=1$: 791 \\
  &&&&&&& $S=2$: 236 \\ \hline
\end{tabular}\\}
\caption{Reduction of the Hilbert space for the
$(S_z,L)=(0,-25)$ subspace of a four-electron system in the
two-Landau-level approximation using CM and total spin as good quantum
numbers. 
\label{tablecm}}
\end{table}
To illustrate the method and compare it with previous work 
\cite{maksym90,yang93,hawrylak93b,palacios94,oaknin96,hawrylak96,wojs97b,reimann99}
we show the results for four electrons in the range of magnetic fields
$0-8$~T for GaAs parameters ($\varepsilon=12.4$, $m^\ast=0.067$), 
confinement energy $\omega_0=3$ meV, and one and two Landau levels.
In Fig.~\ref{figcm} we show the energies of the ground and the two
lowest-lying excited states of each subspace $L$, $S$ with $C_-=0$,
without the Zeeman energy ($g=0$).  
The spectrum in Fig.~\ref{figcm} exhibits an energy band consisting of
ground states of all $S$-subspaces, is separated 
from the excited states by $\Delta E\approx 0.6$ meV (marked by
the double arrow). 
Since the energy needed to construct an excitation of the CM motion is
of the same order of magnitude ($\omega_-=0.62$ meV for $8$ T and higher
for smaller magnetic fields), this separation clearly
shows that the lowest-lying excited states of the quantum dot for
magnetic field between $4$ and $8$ T are all excitations of the
relative motion.

\begin{figure}[h]
\includegraphics*[width=0.9\textwidth]{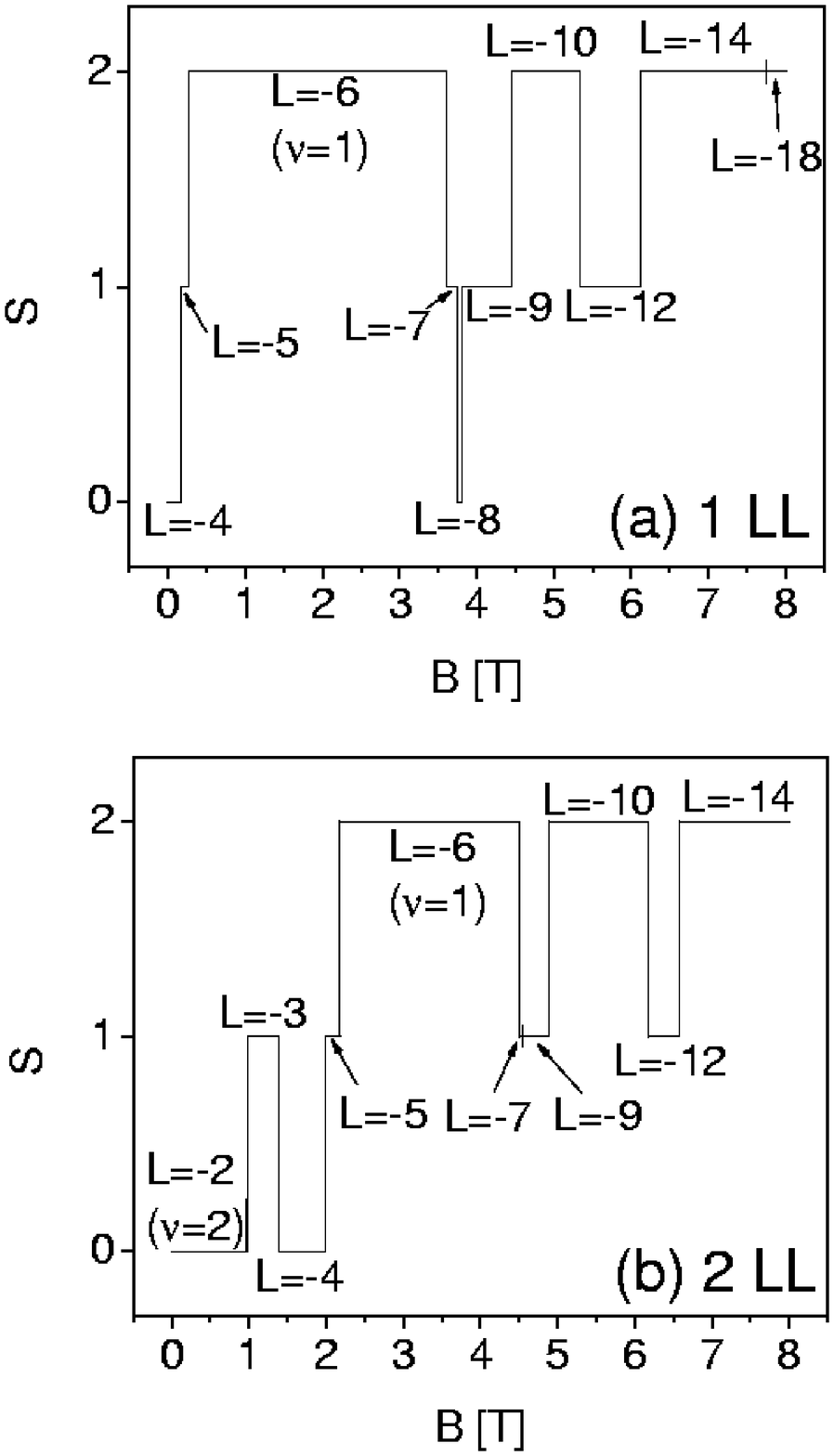}
\caption{Phase diagrams for a four-electron system with confinement energy
$3$ meV and GaAs Zeeman energy ($g=-0.44$) as a function of the
magnetic field from 0 to $8$ T in one-Landau-level (a) and 
two-Landau-level approximation (b).
\label{figcmphase}
}
\end{figure}

In Fig.~\ref{figcmphase} we show
the evolution of the ground state energy with the magnetic field in
the one- and two-Landau-level approximation with GaAs Zeeman energy
(g=-0.44).
In the one-Landau-level approximation we do not find the Hund's rule
state expected at $B=0$ nor the $\nu=2$ spin-singlet droplet with
$S=0$, $L=-2$.
The first quantum Hall state is the spin-polarized state with $S=2$,
$L=-6$. 
 Starting at $B=4$ T we find the reconstruction of the $\nu=1$
droplet. 
This correlation-driven effect is expressed in a number of ground
state transitions with decreasing angular momentum
(Fig.~\ref{figcmphase}(a), (b)). 
The ground states change both angular momentum and spin, in agreement
with previous work
\cite{maksym90,yang93,hawrylak93b,palacios94,oaknin96,hawrylak96,wojs97b,reimann99}.
In Fig.~\ref{figcmphase}(b) we show the evolution of the ground state
in the two-Landau-level approximation.
We still miss the Hund's rule $B=0$ state, but recover the $\nu=2$
state with $L=-6$ and $S=0$.

The qualitative agreement between the lowest- (Fig.~\ref{figcmphase}(a)) 
and two-Landau-level (Fig.~\ref{figcmphase}(b))
approach is good for high magnetic fields, i.e., beyond the $\nu=1$
quantum Hall droplet.
The only difference is the $(S,L)=(0,-8)$ phase which is present
in Fig.~\ref{figcmphase}(a) , and  missing in Fig.~\ref{figcmphase}(b).  
The absence of the low-spin phase is due to the weakening of Coulomb
interactions by inclusion of higher Landau levels.
The second effect is that the fields corresponding to the ground-state
transitions are lower in the one-Landau-level due to the fact that in
this approximation the correlation effects are underestimated.

\begin{figure}[h]
\includegraphics*[width=0.9\textwidth]{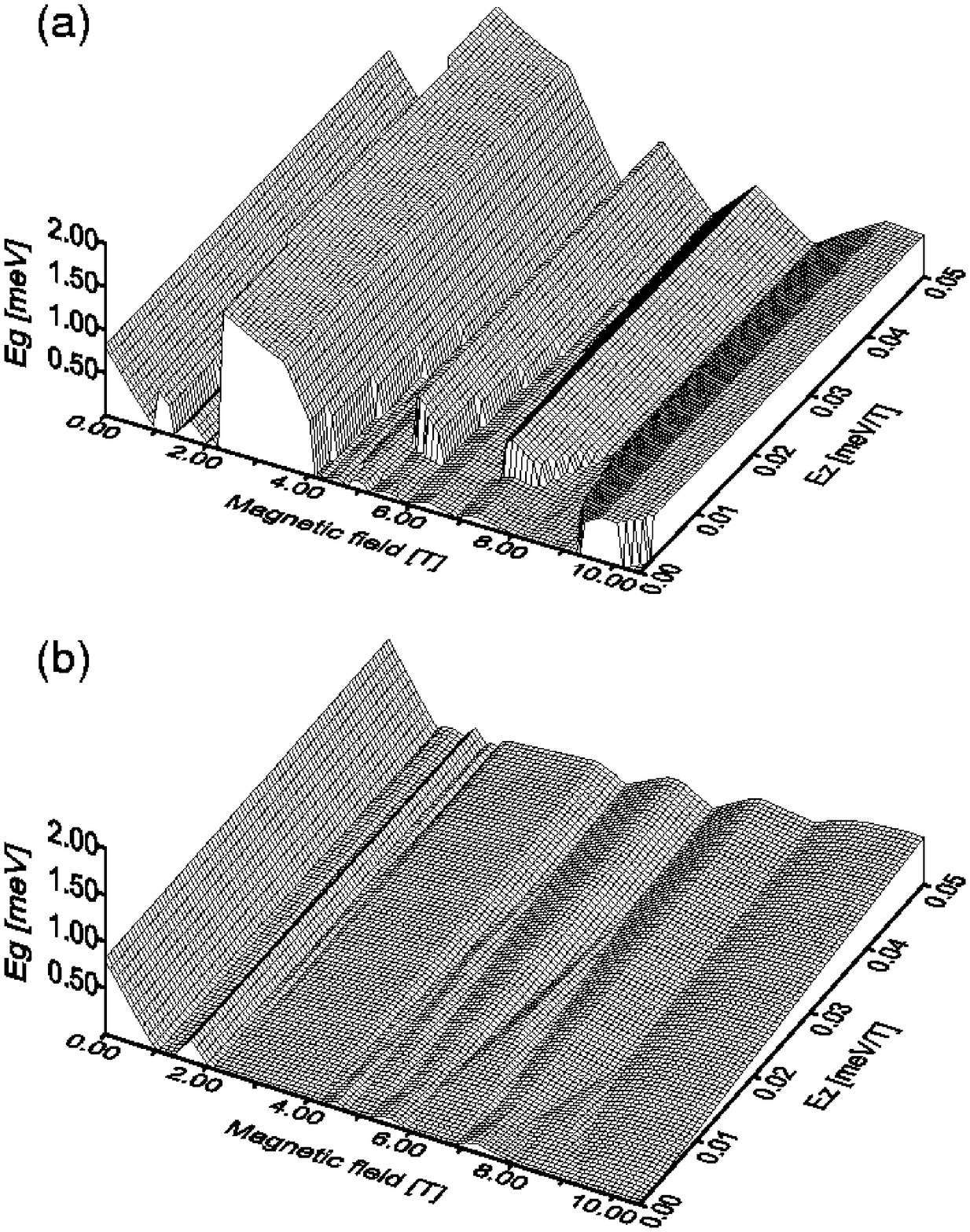}
\caption{Energy gap between the ground and the lowest-lying excited
  state with the same $S_z$ (a) and with $\Delta S_z=\pm 1$ (b)
  for a four-electron system as a function of magnetic field and
  Zeeman energy. 
\label{figexcit}
}
\end{figure}

An advantage of the CI method over QMC and SVM methods is the ability
to calculate excited states. 
In Fig.~\ref{figexcit} we show the energy gap between the ground and
the lowest-lying excited states as a function of the magnetic field and
Zeeman energy for the N=4 electron system, calculated in the
two-Landau-level approximation. 
The magnetic field changes from from $0$ to $10.5$ T,  where the
filling factor $\nu=1/3$-droplet (angular momentum $-18$) emerges.  
In Fig.~\ref{figexcit}(a) we plot the energy gap between the ground
state and the lowest excited state with the same $S_z$ (selection rule
$\Delta S_z=0$, without restrictions for the quantum numbers $L$,
$S$). 
This selection rule applies to  charge excitations. 
Consequently, we cannot expect a systematic increase of the gap as a
function of the Zeeman energy. 
All the structures in the diagram for increasing Zeeman energy are
caused by transitions to higher polarized states. 
In contrast, we used the selection rule $\Delta S_z=\pm 1$ for 
Fig.~\ref{figexcit}(b). 
This kind of excitation should be observed in microwave-radiation or
in spin-flip-Raman experiments. 
The characteristic features in this case are energy gaps linearly growing
with Zeeman energy. 
This behavior is especially pronounced in the regime of the $\nu=1$
droplet around $B=3$ T and in the regime of the $\nu=1/3$ droplet
around $B=10$ T, where both ground and first excited state are stable.

\section{Conclusions}

We presented a configuration interaction method optimized for
Fock-Darwin states of two-dimensional quantum dots with an axially
symmetric parabolic confinement potential subject to perpendicular
magnetic field. 
The optimization explicitly accounts for  geometrical and dynamical
symmetries of Fock-Darwin single-particle states and for 
many-particle symmetries associated with the center-of-mass motion
and with total spin. 
As a key element we introduced a block structure to cluster Slater
determinants which are coupled by $\hat {\bf S}^2$ and/or 
$\hat C_-$. 
By diagonalizing these small blocks we managed to rotate the
many-particle basis set into eigenstates of $\hat {\bf S}^2$ and/or
$\hat C_-$, and used them to calculate the Hamiltonian matrix. 
This reduced the size of the basis set and increased the accuracy of
exact diagonalization calculations for quantum dots.
The results compare well with available quantum Monte-Carlo and 
stochastic-variational results.
Thus we show that the CI method gives reliable results for the ground 
and excited states and allows to study  the effects of electron
correlation and the magnetic field over a broad range of system
parameters. 
It is intended serve as a benchmark for other methods which can be
extended to larger particle numbers.

\section{Acknowledgement}
A.\ W.\ thanks the German Academic Exchange Service (Grant
no.\ D/00/05486), the RRZE Erlangen, and the Institute for
Microstructural Sciences, National Research Council of Canada for
financial support and hospitality.


\begin{thebibliography}{00}


\bibitem{jacak98}
L. Jacak, P. Hawrylak, and A. Wojs, 
{\em Quantum Dots} (Springer, Berlin, 1998).

\bibitem{maksym90}
P.A. Maksym and T. Chakraborty,
Phys. Rev. Lett. {\bf 65}, 108 (1990).

\bibitem{merkt91}
U. Merkt, J. Huser, and M. Wagner,
Phys. Rev. B {\bf 43}, 7320 (1991).

\bibitem{pfannkuche93}
D. Pfannkuche, V. Gudmundsson, and P.A. Maksym, 
Phys. Rev. B {\bf 47}, 2244 (1993).

\bibitem{hawrylak93}
P. Hawrylak and D. Pfannkuche, 
Phys. Rev. Lett. {\bf 70}, 485 (1993).

\bibitem{yang93}
S.-R. Eric Yang, A.H. MacDonald, and M.D. Johnson,
Phys. Rev. Lett. {\bf 71}, 3194 (1993).

\bibitem{hawrylak93b}
P. Hawrylak, Phys. Rev. Lett. {\bf 71}, 3347 (1993).

\bibitem{palacios94}
J.J. Palacios, L. Martin-Moreno, G. Chiappe, E. Louis,
and C. Tejedor,
Phys. Rev. B {\bf 50}, 5760 (1994).

\bibitem{wojs95}
A. Wojs and P. Hawrylak, Phys. Rev. B {\bf 51}, 10880 (1995).

\bibitem{oaknin96}
J.H. Oaknin, L. Martin-Moreno, J.J. Palacios, and 
C. Tejedor, Phys. Rev. Lett. {\bf 74}, 5120 (1995).

\bibitem{wojs96}
A. Wojs and P. Hawrylak, Phys. Rev. B {\bf 53}, 10841 (1996).

\bibitem{maksym96}
P.A. Maksym, Phys. Rev. B {\bf 53}, 10871 (1996).

\bibitem{hawrylak96}
P. Hawrylak, A. Wojs, and J.A. Brum, 
Phys. Rev. B {\bf 54}, 11397 (1996).

\bibitem{wojs97}
A. Wojs and P. Hawrylak, Phys. Rev. B {\bf 55}, 13066 (1997).

\bibitem{wojs97b}
A. Wojs and P. Hawrylak, Phys. Rev. B {\bf 56}, 13227 (1997).

\bibitem{eto97}
M. Eto, Jpn. J. Appl. Phys. {\bf 36}, 3924 (1997).

\bibitem{eto97b}
M. Eto, J. Phys. Soc. Japan {\bf 66}, 2244 (1997).

\bibitem{maksym98}
P.A. Maksym, Physica B {\bf 249-251}, 233 (1998).

\bibitem{imamura98}
H. Imamura, H. Aoki, and P.A. Maksym, 
Phys. Rev. B {\bf 57}, R4257 (1998).

\bibitem{imamura98b}
H. Imamura, H. Aoki, and P.A. Maksym, 
Physica B {\bf 249-251}, 214 (1998).

\bibitem{hawrylak99}
P. Hawrylak, C. Gould, A.S. Sachrajda, Y. Feng, and Z. Wasilewski,
Phys. Rev. B {\bf 59}, 2801 (1999).

\bibitem{creffield99}
C.E. Creffield, W. H{\"a}usler, J.H. Jefferson, and S. Sarkar, 
Phys. Rev. B {\bf 59}, 10719 (1999).

\bibitem{bruce00}
N.A. Bruce and P.A. Maksym, Phys. Rev. B {\bf 61}, 4718 (2000).

\bibitem{reimann00}
S.M. Reimann, M. Koskinen, and M. Manninen, 
Phys. Rev. B {\bf 62}, 8108 (2000).

\bibitem{mikhailov02}
S.A. Mikhailov, Phys. Rev. B {\bf 65}, 115312 (2002).



\bibitem{bolton96}
F. Bolton, Phys. Rev. B {\bf 54}, 4780 (1996).

\bibitem{harju99}
A. Harju, V.A. Sverdlov, R.M. Nieminen, and V. Halonen, 
Phys. Rev. B {\bf 59}, 5622 (1999).

\bibitem{harju99b}
A. Harju, S.  Siljam{\"a}ki, and R.M. Nieminen,
Phys. Rev. B {\bf 60}, 1807 (1999).

\bibitem{harju02a}
A. Harju, S. Siljam{\"a}ki, and R.M. Nieminen,
Phys. Rev. B {\bf 65}, 075309 (2002).

\bibitem{harju02b}
A. Harju, S. Siljam{\"a}ki, R.M. Nieminen, V.A. Sverdlov, and
P. Hyv{\"o}nen, Phys. Rev. B {\bf 65}, 121306 (2002).

\bibitem{egger99}
R. Egger, W. H{\"a}usler, C.H. Mak, and H. Grabert,
Phys. Rev. Lett. {\bf 92}, 3320 (1999).

\bibitem{pederiva00}
F. Pederiva, C.J. Umrigar, and E. Lipparini,
Phys. Rev. B {\bf 62}, 8120 (2000).

\bibitem{pederiva03}
F. Pederiva, C.J. Umrigar, and E. Lipparini,
Phys. Rev. B {\bf 68}, 089901 (2003).

\bibitem{filinov01}
A.V. Filinov, M. Bonitz, and Y.E. Lozovik, 
Phys. Rev. Lett. {\bf 86}, 3851 (2001).


\bibitem{yannouleas99}
C. Yannouleas and U. Landman, Phys. Rev. Lett. {\bf 82}, 5325 (1999).

\bibitem{reusch01}
B. Reusch, W. H{\"a}usler, and H. Grabert, 
Phys. Rev. B {\bf 63}, 113313 (2001).


\bibitem{koskinen97}
M. Koskinen, M. Manninen, and S.M. Reimann, 
Phys. Rev. Lett. {\bf 79}, 1389 (1997).

\bibitem{austing99}
D.G. Austing, S. Sasaki, S. Tarucha, S.M. Reimann, M. Koskinen, 
and M. Manninen, Phys. Rev. B {\bf 60}, 11514 (1999).

\bibitem{hirose99}
K. Hirose and N.S. Wingreen, Phys. Rev. B {\bf 59}, 4604 (1999).

\bibitem{steffens99}
O. Steffens and M. Suhrke, Phys. Rev. Lett. {\bf 82}, 3891 (1999).

\bibitem{wensauer00}
A. Wensauer, O. Steffens, M. Suhrke, and U. R\"ossler, 
Phys. Rev. B {\bf 62}, 2605 (2000).

\bibitem{wensauer01}
A. Wensauer, J. Kainz, M. Suhrke, and U. R\"ossler, 
phys. stat. sol. (b) {\bf 224}, 675 (2001).


\bibitem{varga01}
K. Varga, P. Navratil, J. Usukura, and Y. Suzuki, 
Phys. Rev. B {\bf 63}, 205308 (2001).


\bibitem{fock28}
V. Fock, Z. Phys. {\bf 47}, 446 (1928).

\bibitem{darwin31}
C.G. Darwin, Proc. Cambridge Phil. Soc. {\bf 27}, 86 (1931).


\bibitem{hawrylak_sc93}
P. Hawrylak, Solid State Commun. {\bf 88}, 475 (1993).



\bibitem{capelle02}
K. Capelle and G. Vignale, Phys. Rev. B {\bf 65}, 113106 (2002).







\bibitem{chakra92}
T. Chakraborty, Comments Conde. Matt. Phys. {\bf 16}, 35 (1992).



\bibitem{reimann99}
S.M. Reimann, M. Koskinen, M. Manninen, and B.R. Mottelson,
Phys. Rev. Lett. {\bf 83}, 3270 (1999).


\end{thebibliography}
\end{document}